\documentclass{PoS}
\usepackage{subfigure,amsmath}
\title{MCRG Minimal Walking Technicolor}

\ShortTitle{MCRG Minimal Walking Technicolor}

\author{Simon Catterall\\
        Syracuse University\\
        E-mail: \email{smc@physics.syr.edu}}

\author{Luigi Del Debbio\\
        Edinburgh University\\
        E-mail: \email{luigi.del.debbio@ed.ac.uk}}

\author{Joel Giedt\\
        Rensselaer Polytechnic Institute\\
        E-mail: \email{giedtj@rpi.edu}}

\author{\speaker{Liam Keegan}\\
        Edinburgh University\\
        E-mail: \email{liam.keegan@ed.ac.uk}}

\abstract{We present a Monte Carlo renormalisation group study of the SU(2) gauge theory with two Dirac fermions in the adjoint representation. Using the two lattice matching technique recently advocated and exploited in Ref.~\cite{Hasenfratz:2009ea}, we measure the running of the coupling and the anomalous mass dimension.}

\FullConference{The XXVIII International Symposium on Lattice Field Theory, Lattice2010\\
		June 14-19, 2010\\
		Villasimius, Italy}

\begin{document}

\section{Introduction}

Technicolor theories with fermions in higher representations of the gauge group can potentially provide a dynamical electroweak symmetry breaking mechanism without conflicting with electroweak precision data. Minimal Walking Technicolor is an example of such a theory, a SU(2) gauge theory with two Dirac fermions in the adjoint representation~\cite{Sannino:2004qp,Luty:2004ye}. It is expected from perturbation theory to be in or near to the conformal window, although since this is at relatively strong coupling non-perturbative results are desirable. Lattice simulations have indeed found that the gauge coupling runs very slowly~\cite{Bursa:2009we,DelDebbio:2009fd,Hietanen:2009az}, more slowly than the perturbative prediction, although distinguishing between conformal and near-conformal behaviour is an inherently difficult task. In order to be phenomenologically viable the theory must have a large anomalous dimension $(\gamma\sim1)$~\cite{Holdom:1981rm,Yamawaki:1985zg,Appelquist:1986an}. The conjectured all-order beta function predicts $\gamma=0.75$~\cite{Ryttov:2007cx}, and the anomalous dimension has been measured non-perturbatively in some recent lattice studies~\cite{Bursa:2009we,DelDebbio:2010hx}, which have all found lower values.

In this work we measure the anomalous mass dimension using the Monte Carlo Renormalisation Group, a technique which has recently been used to investigate theories with many flavours of fermions in the fundamental representation~\cite{Hasenfratz:2009ea,Hasenfratz:2010fi}. We consider the evolution of all possible couplings of the system under Wilson RG block transformations, where with each blocking step ultraviolet fluctuations are integrated out. Fixed points are characterised by the number of relevant couplings, which have positive scaling dimensions and flow away from the fixed point. Irrelevant couplings have negative scaling dimensions, and flow towards the fixed point, so that their IR values are independent of their UV values.

\section{Method}

With each RG step, changing the scale by a factor $s$, irrelevant couplings will flow towards the fixed point, and relevant couplings will flow away from it. After a few steps the irrelevant couplings should die out, leaving the flow following the unique renormalised trajectory (RT). If we can identify two sets of couplings which end up at the same point along the RT after the same number of steps, then they must have the same lattice correlation lengths, $\hat{\xi}=\hat{\xi}'$. Since the physical correlation length $\xi = \hat{\xi} a$ should not be changed by the RG transform, this means that they both must have the same lattice spacing $a$, or inverse cutoff $\Lambda^{-1} \sim a$. If they end up at the same point, but one takes an extra step, then their lattice correlation lengths must differ by a factor $s$, and hence so must their UV cutoffs.

To identify such a pair of couplings, we need to show that after $n$ and $(n-1)$ RG steps respectively their actions are identical. Explicitly calculating the actions would be complicated, but instead the gauge configurations themselves can be RG block transformed, and showing that the expectation values of all observables on these gauge configurations agree is equivalent to directly comparing the actions that generated them.

\subsection{2-Lattice Matching Procedure}

Starting with the SU(2) pure gauge theory, where the gauge coupling is the only revelant parameter, the procedure is as follows.

\begin{enumerate}
 \item Generate an ensemble of gauge configurations with an action $S(\beta)$ on a $L^4$ lattice.
 \item Block these $n$ times to produce an ensemble of configurations on a $(L/s^n)^4$ lattice, and measure the expectation values of various observables on them.
 \item Generate a new ensemble of gauge configurations with an action $S(\beta')$ on a $(L/s)^4$ lattice, for a range of values of $\beta'$.
 \item Block each of these $n-1$ times to produce an ensemble of configurations on a $(L/s^n)^4$ lattice, and measure the same observables for each $\beta'$.
 \item Interpolate in $\beta'$ such that each observable after taking $n$ steps on the larger lattice agrees with the same observable after taking $(n-1)$ steps on the smaller lattice.
 \item Repeat for different $n$, e.g. for $s=2,L=32$, three values can be used: $n=2,3,4$.
\end{enumerate}

We have now identified, for each $n$, a pair of bare gauge couplings $(\beta,\beta')$, with lattice correlation lengths that differ by a factor $s$, $\hat{\xi}' = \hat{\xi}/s$. In the limit $n \rightarrow \infty$, the quantity $\Delta \beta = \beta - \beta' \equiv s_{b}(\beta;s)$ is the step scaling function for the bare gauge coupling. This is the analog of the Schr\"odinger Functional step scaling function for the renormalised coupling, $\sigma(u,s)$, and in the UV limit where $\overline{g}^2 \rightarrow g_0^2 = 2N/\beta$, there is a simple relation between the two:
\begin{equation}
\frac{s_{b}(\beta;s)}{\beta} = \frac{\sigma(u,s)}{u} - 1
\end{equation}

We use the following $s=2$ RG blocking transform~\cite{Hasenfratz:2009ea}
\begin{equation}
V_{n,\mu} = Proj\left[(1-\alpha)U_{n,\mu}U_{n+\mu,\mu}+\frac{\alpha}{6}\sum_{\nu\neq\mu}U_{n,\nu}U_{n+\nu,\mu}U_{n+\mu+\nu,\mu}U_{n+2\mu,\nu}^{\dagger}\right]
\end{equation}

Here $\alpha$ is a free parameter, which can be varied to optimise the transformation. Changing $\alpha$ changes the location of the fixed point, and how quickly the unique RT is approached in a given number of steps. Ideally it should be chosen such that

\begin{itemize}
 \item All operators predict the same matching coupling between $(n,n-1)$ pairs for a given blocking step $n$. (Deviations are a measure of the systematic error from not being at exactly the same point along the RT) 
 \item Consecutive blocking steps predict the same matching coupling, i.e. the coupling for which $(n,n-1)$ pairs agree should be the same for all $n$. (Deviations show that the RT is still being approached in the irrelevant directions)
\end{itemize}

\section{Pure Gauge Results}

\begin{figure}[ht]
  \centering
\subfigure[Plaquette Matching]{\label{fig:PURE_plaq_a}\includegraphics[angle=270,width=7.5cm]{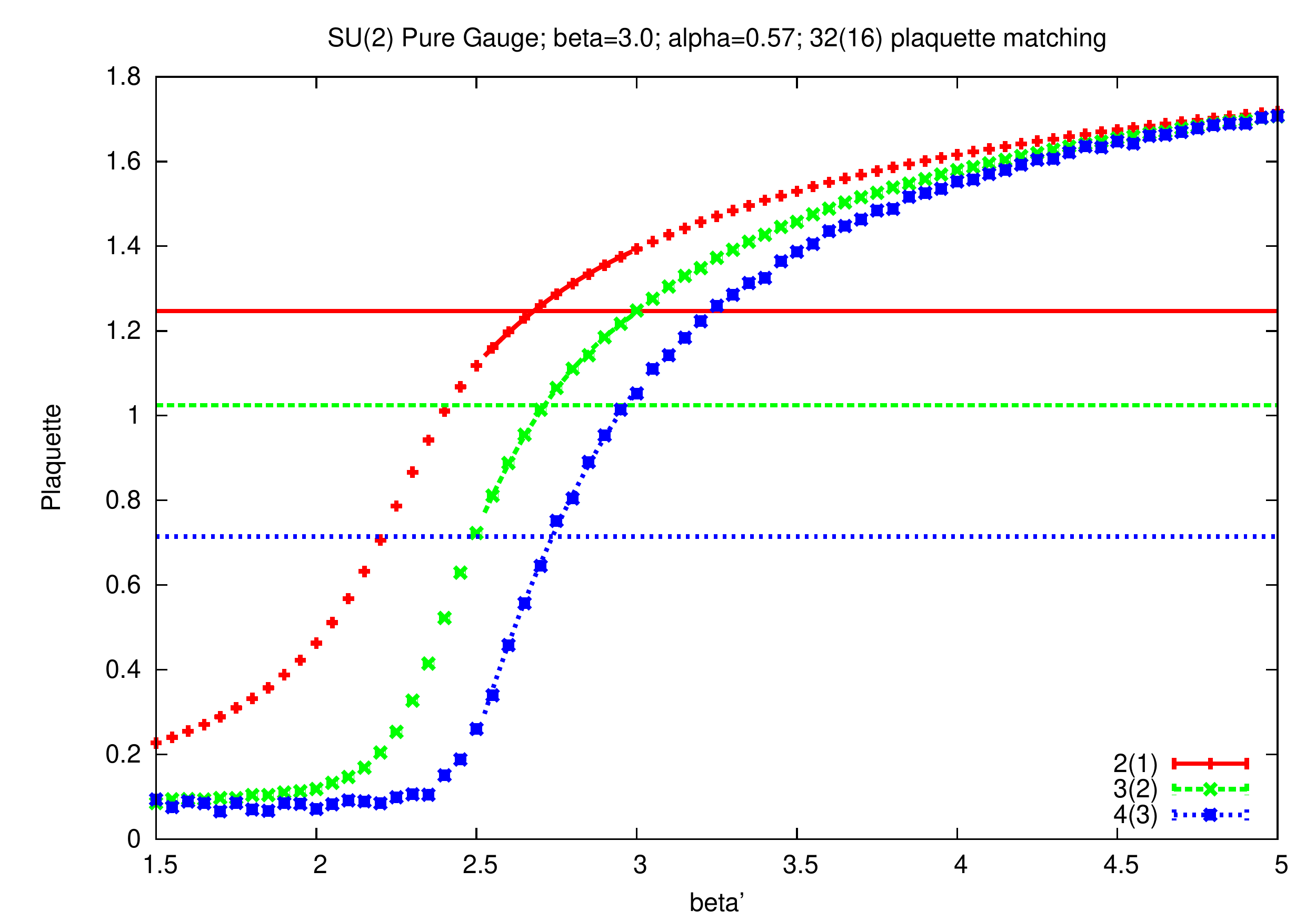}}
\subfigure[$\alpha$-Optimisation]{\label{fig:PURE_plaq_b}\includegraphics[angle=270,width=7.5cm]{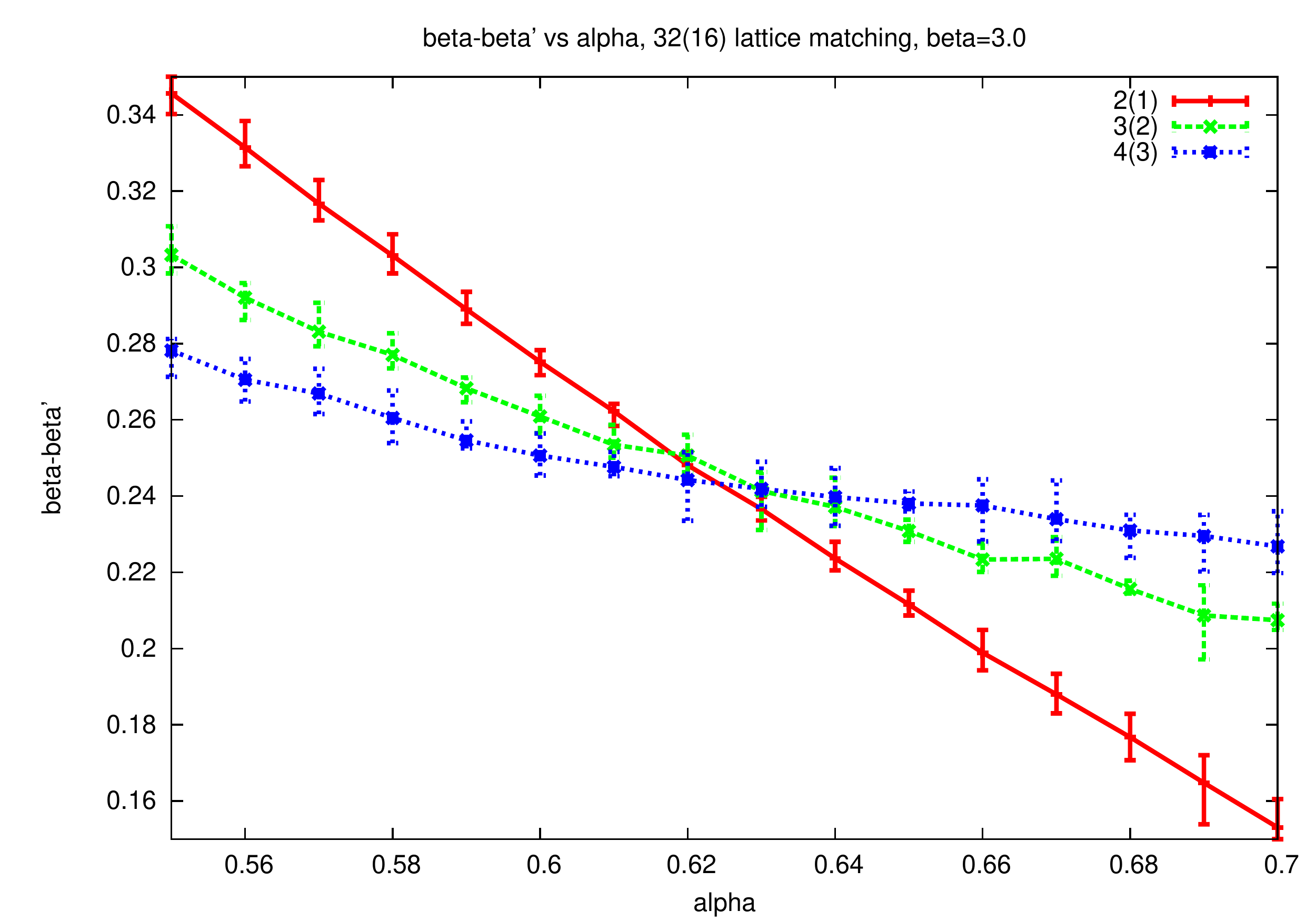}}
  \caption{An example of the matching of the plaquette in $\beta$ for the pure gauge case. This is repeated for each observable to give a systematic error for each matching, then $\alpha$ is varied such that all blocking steps predict the same matching.}
  \label{fig:PURE_plaq}
\end{figure}

As an initial test of the method we determined the step scaling of the bare coupling for the SU(2) pure gauge theory.  We matched in $\beta$, using $200$ SU(2) pure gauge configurations for each $\beta$ on $32^4$ and $16^4$ lattices. We matched in the plaquette and in the three six-link loops. An example of the matching of the plaquette is shown in Figure~\ref{fig:PURE_plaq_a}. The red, green and blue horizontal lines show the average plaquette on the $32^4$ lattice after 2, 3 and 4 blocking steps respectively, at $\beta=3.0, \alpha=0.57$. The interpolated red, green and blue points show the average plaquette on the $16^4$ lattice after 1, 2 and 3 blocking steps respectively, as a function of $\beta'$, also at $\alpha=0.57$. The value of $\beta'$ where the two red lines intersect gives the matching coupling for $n=2$, similarly the green and blue lines give the matching coupling for $n=3$ and $4$.

\begin{figure}[ht]
  \centering
\includegraphics[angle=0,width=11.5cm]{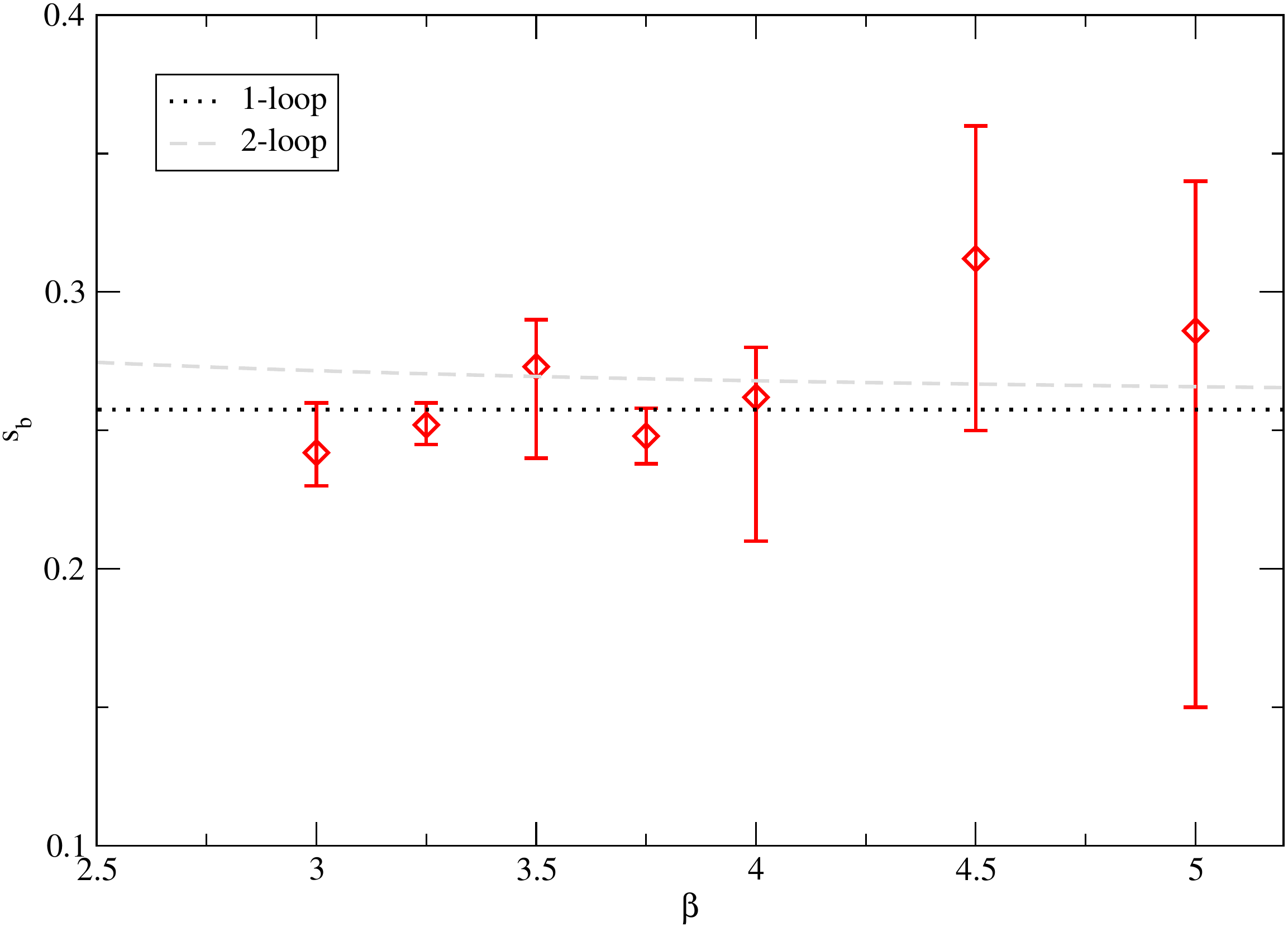}
  \caption{Bare step scaling $s_{b}(\beta;s=2)$ for the pure gauge theory.}
  \label{fig:PURE_sb}
\end{figure}

This matching is repeated for each observable, and the spread of predicted matchings for each $n$ gives a systematic error on the central matching value. The whole procedure is then repeated for various values of $\alpha$, as shown in Figure~\ref{fig:PURE_plaq_b}. In the limit $n \rightarrow \infty$ the predicted matching value should converge on a single value that does not change if $n$ is increased, and is independent of $\alpha$. With a finite number of steps we can exploit the $\alpha$-dependence to find an optimal value of $\alpha$, where subsequent RG steps predict the same matching value. The intersection of the last two blocking steps gives a central value for $s_{b}(\beta=3.0;s=2)$, while the range of couplings for which any of the blocking steps intersect within errors gives the uncertainty on this central value. This was repeated for other values of $\beta$, and the resulting step scaling of the bare coupling is shown in Figure~\ref{fig:PURE_sb}. It agrees well with the perturbative prediction.

\section{Anomalous Dimension Results}

At an IRFP in the full theory with dynamical fermions, the gauge coupling is expected to be irrelevant, leaving the mass as the only relevant operator. This means we should in principle be able to match in the mass at arbitrary couplings, as long as we have sufficient RG steps for the gauge coupling to flow to its fixed point value. We match observables as for the pure gauge case, but instead of matching in $\beta$, we set $\beta'=\beta$, and match pairs of bare masses $(m,m')$.

\begin{figure}[ht]
  \centering
\subfigure[Plaquette Matching]{\includegraphics[angle=270,width=7.5cm]{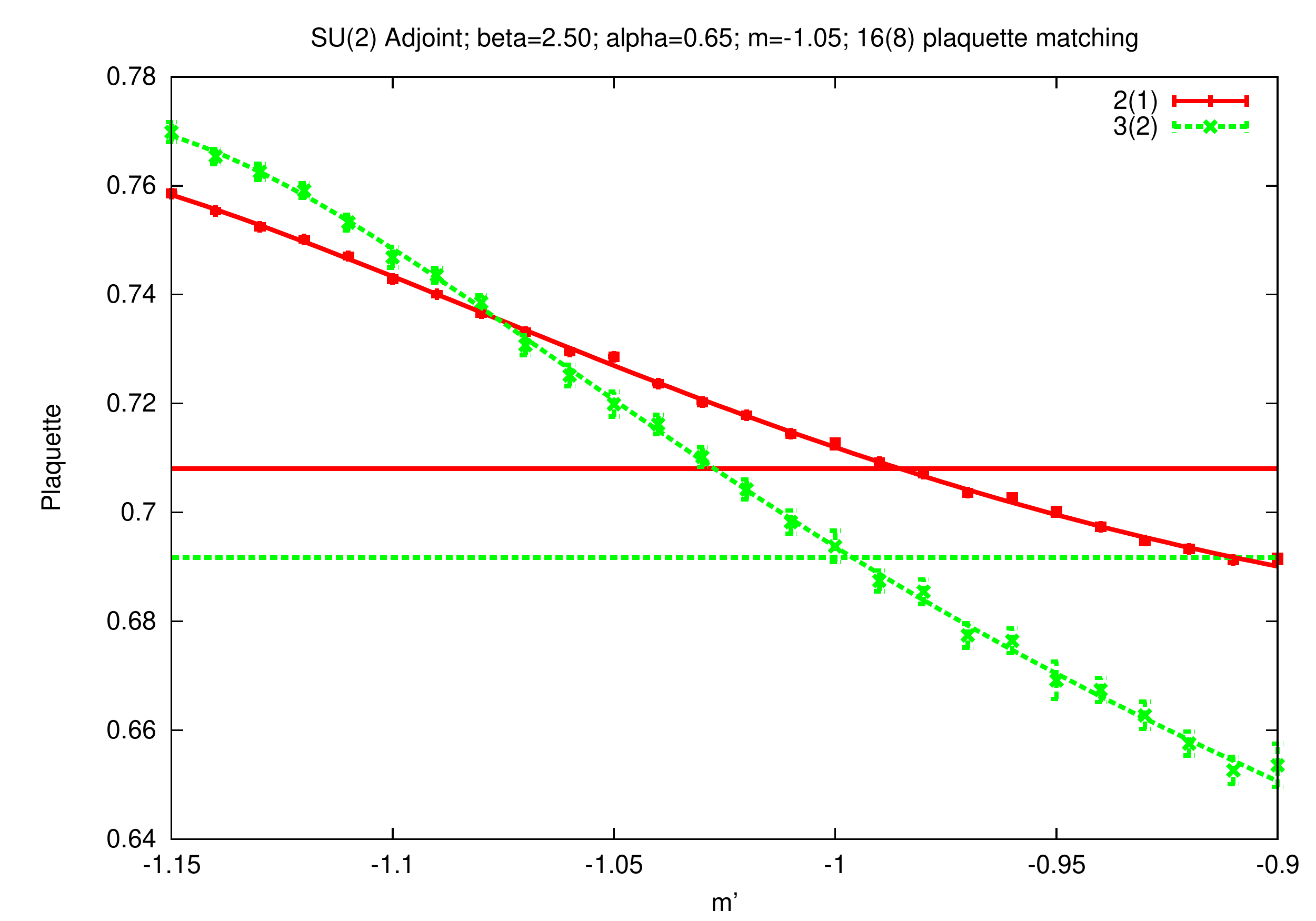}}
\subfigure[$\alpha$-Optimisation]{\includegraphics[angle=270,width=7.5cm]{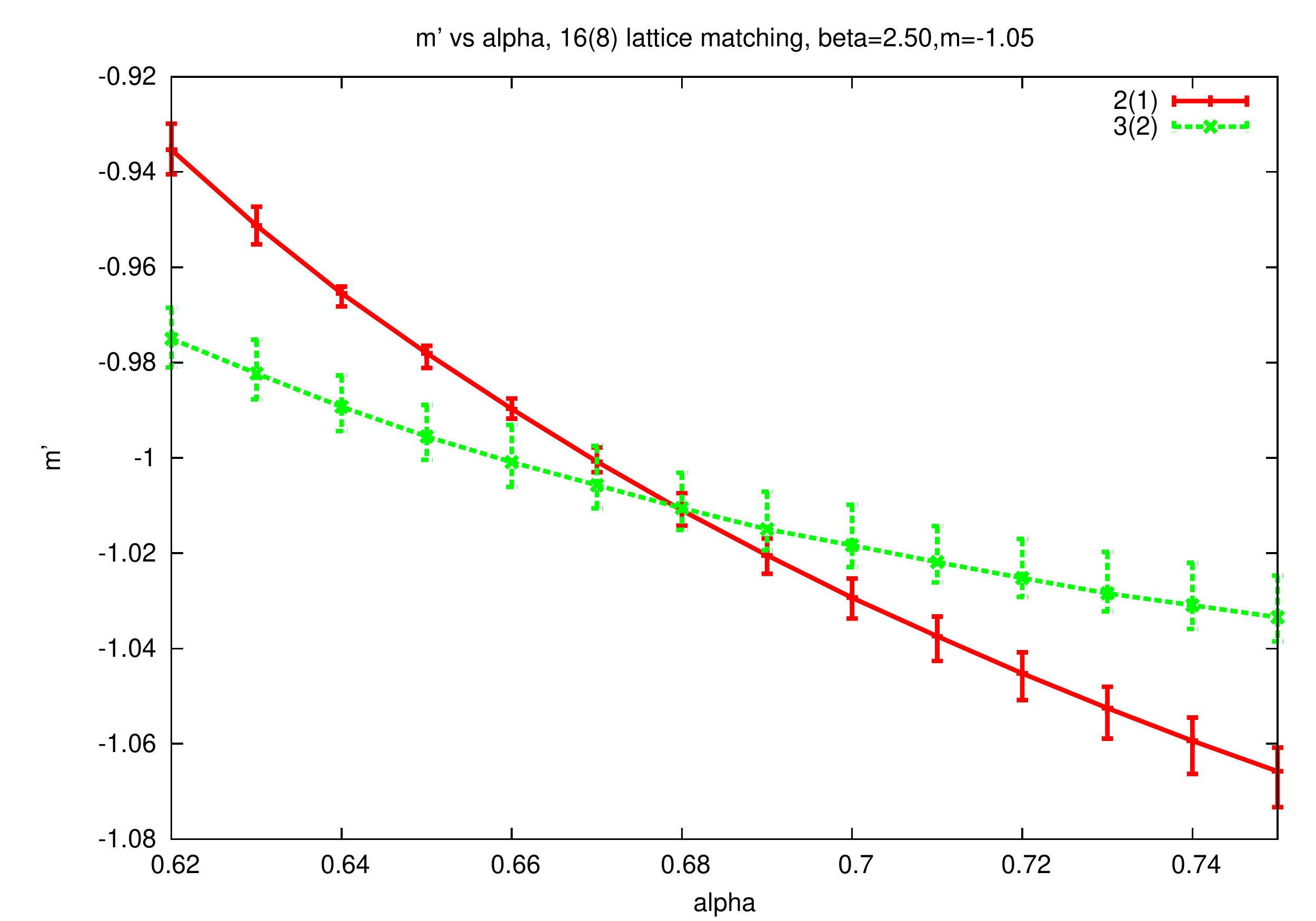}}
  \caption{An example of the matching of the plaquette in the bare mass $m$. This is repeated for each observable to give a systematic error for each matching, then $\alpha$ is varied such that all blocking steps predict the same matching.}
  \label{fig:MASS_plaq}
\end{figure}

We use the HiRep RHMC code~\cite{DelDebbio:2008zf} with a Wilson plaquette gauge action and adjoint Wilson fermions. We generated $\sim1000$ configurations on $16^4$ and $8^4$ lattices, for a range of bare masses at each $\beta$. This allows two matching steps, after $2(1)$ and $3(2)$ steps on the $16^4(8^4)$ lattices. An example of this matching and subsequent $\alpha$-optimisation is shown in Figure~\ref{fig:MASS_plaq}.

\begin{figure}[ht]
  \centering
\subfigure[$L=8$]{\includegraphics[angle=270,width=7.5cm]{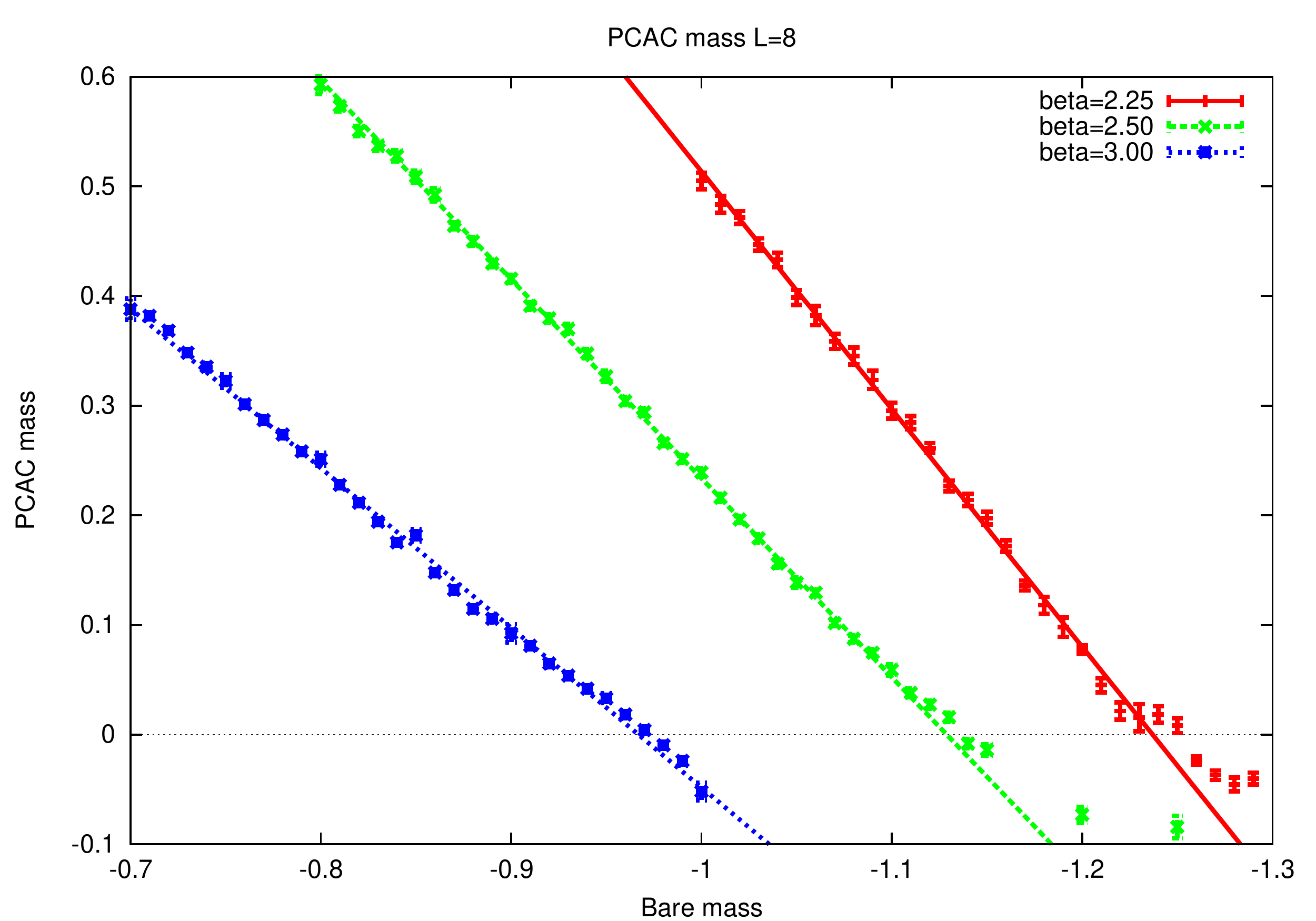}}
\subfigure[$L=16$]{\includegraphics[angle=270,width=7.5cm]{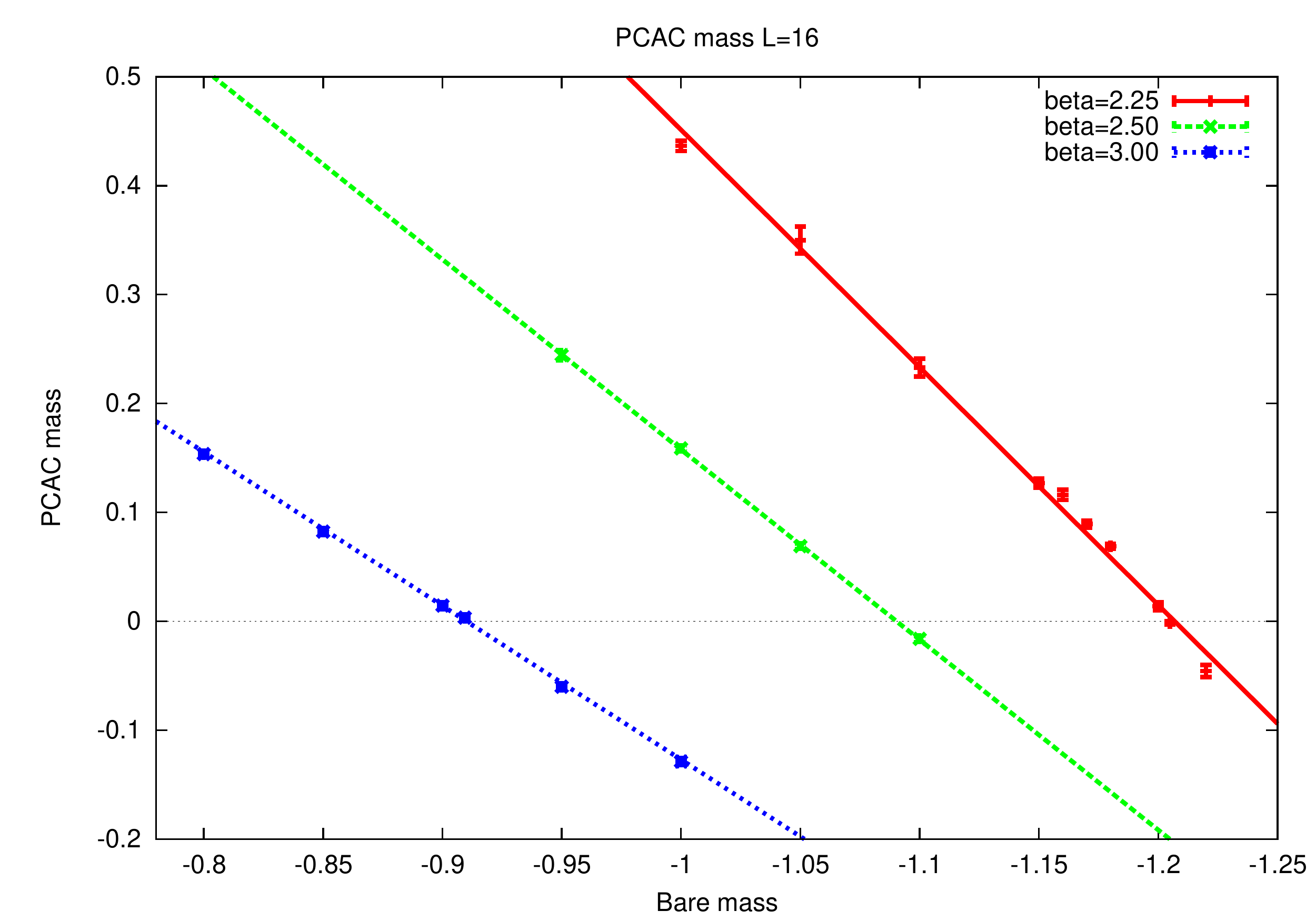}}
  \caption{PCAC mass as a function of the bare mass on $8^4$ and $16^4$ lattices for $\beta=2.25,2.50,3.00$}
  \label{fig:MASS_pcac}
\end{figure}

The bare mass is additively renormalised, so we convert the bare masses to PCAC masses. We measure the PCAC mass averaged over the two central time slices, $\overline{m}$, as a function of bare mass for each $\beta,L$, as shown in Figure~\ref{fig:MASS_pcac}, and use this to convert the pairs of bare masses to pairs of PCAC masses. 
\begin{equation}
\overline{m} = \tfrac{1}{2}\left[m_{PCAC}\left(\tfrac{L}{2}-1\right)+m_{PCAC}\left(\tfrac{L}{2}\right)\right], \quad m_{PCAC}(x_0) = \frac{\tfrac{1}{2}(\partial_0+\partial^{*}_0)f_{A}(x_0)}{2f_{P}(x_0)}
\end{equation}

\begin{figure}[ht]
  \centering
\includegraphics[angle=0,width=12.5cm]{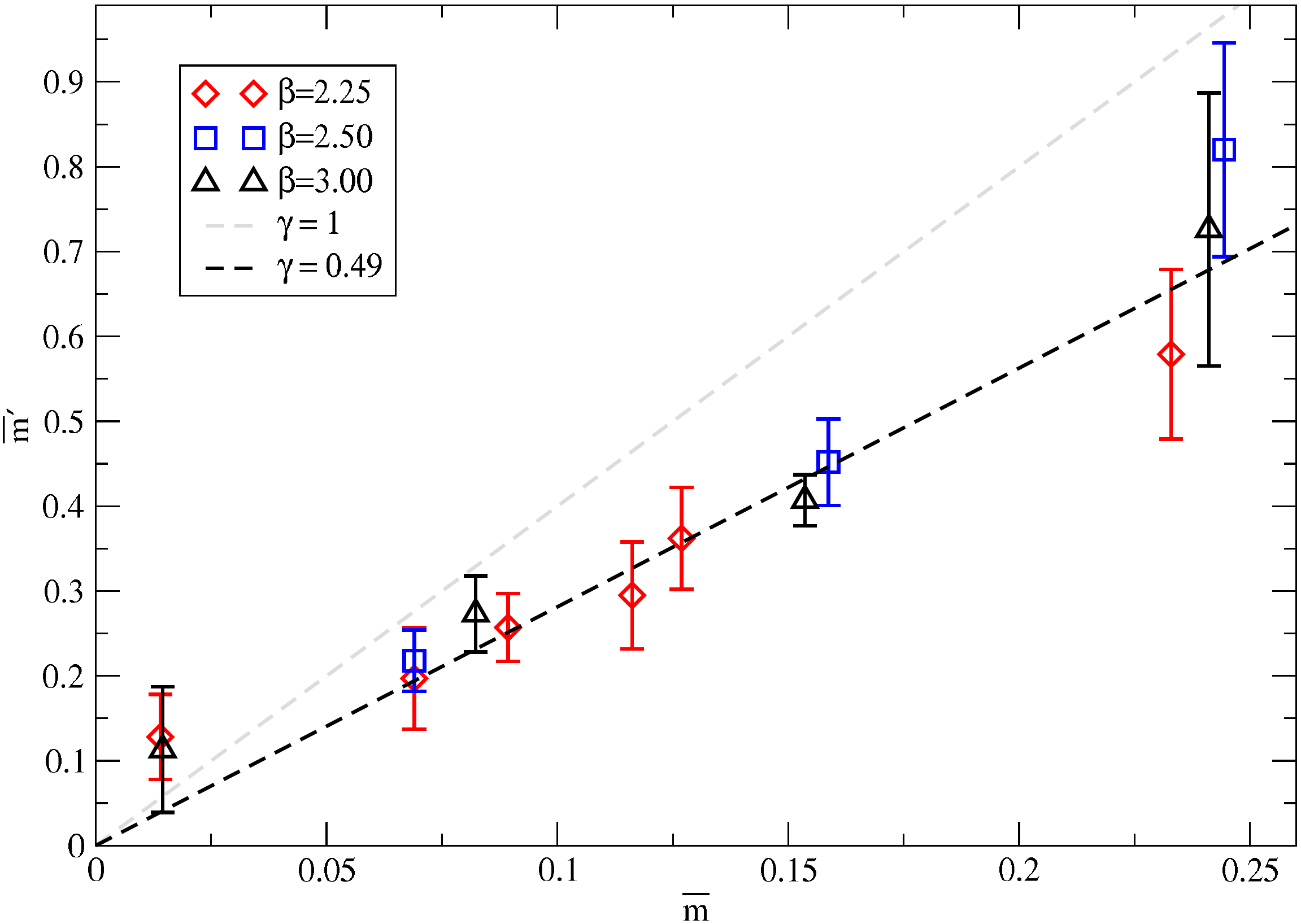}
  \caption{Matching mass pairs. Linear fit gives $\gamma=0.49(13)$. Also shown is the linear fit given by assuming $\gamma=1$, such a large anomalous dimension is clearly excluded by our data.}
  \label{fig:MASS_mass}
\end{figure}

The anomalous mass dimension can then be extracted from the relation $\overline{m}'/\overline{m} = 2^{\gamma+1}$. If we are at an IRFP then this is a scheme-independent quantity, so we should be able to repeat the measurement at any value of $\beta$ and find the same anomalous dimension. We used three values of $\beta$, $\beta=2.25,2.50,3.00$, and all give consistent predictions for the anomalous dimension. The matching PCAC mass pairs and a linear fit are shown in Figure~\ref{fig:MASS_mass}, and predict $\gamma=0.49(13)$.

\section{Coupling}
It would also be interesting to measure the step scaling of the bare coupling $s_{b}(\beta;s)$ for the full theory with dynamical fermions, as was done for the pure gauge case. A fixed point would be indicated by a change of sign in this quantity as the bare coupling is varied from weak to strong coupling. The difficulty is that the mass is a relevant operator, while the coupling is expected to be at best nearly marginal, so that in order for the MCRG to pick out the behaviour of the coupling the mass would have to be tuned to zero.
Furthermore, even if the mass is tuned sufficiently close to zero that we are initially following the evolution of the gauge coupling (assuming it is the least irrelevant remaining operator), we can no longer take the limit $n\rightarrow \infty$, because the coupling will flow to its fixed point value, and the flow of the mass will eventually dominate.

The method would therefore be to tune the mass close to zero initially, then take a few RG steps where the flow is following the gauge coupling, and extract the running of the coupling before the flow in the mass becomes significant. This has been successfully performed in Ref.~\cite{Hasenfratz:2010fi} using staggered fermions for SU(3) gauge theories with 8 and 12 flavours of fundamental fermions.

\section{Conclusion}
We measure the anomalous mass dimension for Minimal Walking Technicolor and find $\gamma=0.49(13)$, in line with previous lattice determinations, and smaller than would be desired for phenomenology. A large anomalous dimension $\gamma=1$ is strongly excluded by our data, with a $\chi^2/d.o.f$ of 8.7 with $12$ degrees of freedom.

We intend to repeat this calculation on larger lattices to reduce our systematic errors, and with more observables to improve our determination of the systematic errors themselves, as well as with different RG blocking transforms to check for universality, and to allow us to go to stronger coupling. We are also planning to match in $\beta$ with all bare masses tuned such that the PCAC masses are as close as possible to zero, in order to look for a fixed point by directly measuring the running of the coupling.

\end{document}